\documentclass{article}
\usepackage{spconf,amsmath,graphicx}

\usepackage{enumitem}
\setlist{nosep, leftmargin=14pt}

\renewcommand{\vec}[1]{\boldsymbol{#1}}

\DeclareMathOperator{\dir}{dir}

\usepackage{calc,tikz}
\usetikzlibrary{arrows.meta}
\usetikzlibrary{calc}
\usetikzlibrary{positioning}
\usetikzlibrary{decorations.markings}
\usetikzlibrary{backgrounds}
\usetikzlibrary{shapes.misc}
\usepackage{stackrel}

\definecolor{softblue}{HTML}{1f77b4}
\definecolor{softred}{HTML}{d62728}

\usepackage[hidelinks]{hyperref}

\title{A Spatiotemporal Illumination Model for \\ 3D Image Fusion in Optical Coherence Tomography}
\name{
\begin{tabular}{c}
    Stefan Ploner$^{1,2}$, Jungeun Won$^{2}$, Julia Schottenhamml$^{1}$,
    Jessica Girgis$^{3}$,
    \\ Kenneth Lam$^{3}$,
    Nadia Waheed$^3$,
    James Fujimoto$^{2}$, Andreas Maier$^{1}$
\end{tabular}
}
\address{$^1$ Pattern Recognition Lab, Friedrich-Alexander-Universität Erlangen-Nürnberg, Erlangen, Germany \\
$^2$ Research Laboratory of Electronics, Massachusetts Institute of Technology, Cambridge, MA, USA \\
$^3$ Department of Ophthalmology, New England Eye Center, Boston, MA, USA
}
\tikzset{
	letter/.style 2 args = {
		append after command={(\tikzlastnode.north west) node[anchor=north west,text=#2,inner sep=1pt,inner xsep=2pt](){\strut #1}} 
	},
	filledletter/.style n args={4}{
		append after command={(\tikzlastnode.north west) node[anchor=north west,text=#2,fill=#3,fill opacity=#4,text opacity=1,inner sep=1pt,inner xsep=2pt](){\strut #1}}
	},
	letter/.default={a}{black},
	filledletter/.default={a}{black}{white}{1}
}
\begin{document}
%
\maketitle
\begin{abstract}
Optical coherence tomography (OCT) is a non-invasive, micrometer-scale imaging modality that has become a clinical standard in ophthalmology.
By raster-scanning the retina, sequential cross-sectional image slices are acquired to generate volumetric data.
In-vivo imaging suffers from discontinuities between slices that show up as motion and illumination artifacts.
We present a new illumination model that exploits continuity in orthogonally raster-scanned
volume data to accurately compensate illumination artifacts.
Our novel spatiotemporal parametrization adheres to illumination continuity both temporally, along the imaged slices, as well as spatially, in the transverse directions.
Yet,
our formulation does not make inter-slice assumptions, which could have discontinuities.
This is the first optimization of a 3D inverse model in an image reconstruction context in OCT.
Evaluation in 68 volumes from eyes with pathology showed reduction of illumination artifacts in 88\% of the data, and only 6\% showed moderate residual illumination artifacts.
The method enables the use of forward-warped motion corrected data, which is more accurate, and enables supersampling and advanced 3D image reconstruction in OCT.
\end{abstract}
%
%
\section{Introduction \& problem statement}
\label{sec:intro}

Optical Coherence Tomography (OCT) is a non-invasive 3D optical imaging modality and standard of care in ophthalmology \cite{Huang1991,Fujimoto2016}, where it is used for retinal imaging.
In OCT angiography (OCTA), scans are repeated and, besides the standard OCT signal, a metric of its variance is computed \cite{Spaide2018}. This displays vasculature, because moving blood cells cause varying backscattering, while it is constant for tissue.
In eyes with pathology, a recent study found severe image artifacts that are associated with the reliability of quantitative biomarkers in 53.5\% of the data \cite{Holmen2020}.
In this study, motion-related artifacts (16.0\%) were among the 3 most prevalent types of artifacts.
Motion artifacts arise because OCT images are acquired by raster-scanning a laser beam across the retinal surface for a multi-second acquisition duration and apply to OCT and OCTA data in the same way.
In their review on image artifacts in OCTA, Anvari et al.\ list 6 types of scanning-related artifacts \cite{Anvari2021}, which we categorize in dominantly distortion-related (displacement, ghosting/doubling, stretch and quilting/crisscross) and illumination-related (blink and banding). Again, illumination-related artifacts exist in OCT and OCTA data, and originate from signal variation in the OCT signal.
A particularity of OCT illumination artifacts is that they can be highly anisotropic due to raster scanning, with slow variation along a scan line, but sharp changes between them (\figurename~1a).

Besides eye tracking-based approaches, image distortions are compensated via retrospective software motion correction methods by aligning multiple scans \cite{Brea2019, Ploner2021}, which can subsequently be merged to improve image quality.
The combined approach was shown to be superior to tracking-based approaches alone \cite{Camino2016}, which are fundamentally limited by latency.
Recently, we demonstrated forward warping motion correction in OCT \cite{Ploner2022}, which achieves a general improvement in accuracy, reducing displacement and ghosting/doubling artifacts. More specifically, forward warping correctly localizes gaps, which solves stretch artifacts, and is able to correctly model eye motion in the slowly-scanned direction, which enables correction of distortions that were responsible for quilting artifacts in prior methods.
Despite the clear need for forward warping motion correction, its utilization is currently hindered in clinical data because it introduces a new challenge for data fusion. It results from illumination differences between the coregistered scans, which is closely related to the second category of scanning artifacts.

Illumination / signal level can vary during acquisition of OCT volumes for various reasons,
including:
Eye blinks;
attenuation when the beam passes through ``floaters'', which are non-stationary opacities in the vitreous;
attenuation from corneal drying;
vignetting on the iris caused by insufficient instrument alignment and eye motion;
and depth-dependent attenuation from intensity roll-off.
Many of these effects worsen with age and in pathology.

Eye blinks block the imaging beam entirely, which reduces signal levels to background noise and can not be corrected.
All other forms of change in illumination are spatially continuous and, normally, vary only slowly in time, which is tolerated in OCT imaging.
Banding artifacts emerge when the temporal continuity is interrupted by scan interruption (which can occur when scanners reset the position in eye tracking), or when the spatial continuity is interrupted between successive B-scans due to eye motion.
\begin{figure}[tbh]
    \centering
    \begin{tikzpicture}[inner sep=0, outer sep=0,baseline=(img.south)]
        \newlength{\picwidth}
        \setlength{\picwidth}{.47\linewidth}
        \node[anchor=south west,filledletter= {\large a}{yellow}{black}{0.5}](img){\includegraphics[keepaspectratio,width=\picwidth]{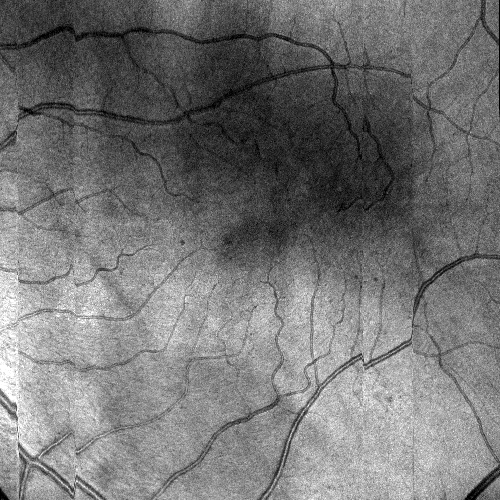}};
        \draw[green!70!black,ultra thick,loosely dashed,rotate=50](.9\picwidth,0\picwidth) ellipse (1.2cm and 1.4cm);
        \draw[<-,>=Latex,ultra thick,softred] (.04\picwidth,.65\picwidth) -- ++(0,.2cm);
        \draw[<-,>=Latex,ultra thick,softred] (.04\picwidth,.05\picwidth) -- ++(0,-.2cm);
        \draw[<-,>=Latex,ultra thick,softred] (.18\picwidth,.9\picwidth) -- ++(0,.2cm);
        \draw[<-,>=Latex,ultra thick,softred] (.18\picwidth,.55\picwidth) -- ++(0,-.2cm);
        \draw[<-,>=Latex,ultra thick,softred] (.35\picwidth,.8\picwidth) -- ++(0,-.2cm);
        \draw[<-,>=Latex,ultra thick,softred] (.54\picwidth,.8\picwidth) -- ++(0,-.2cm);
        \draw[<-,>=Latex,ultra thick,softred] (.82\picwidth,.9\picwidth) -- ++(0,.2cm);
        \draw[<-,>=Latex,ultra thick,softred] (.82\picwidth,.5\picwidth) -- ++(0,-.2cm);
    \end{tikzpicture}
    \hfill
    \begin{tikzpicture}[inner sep=0, outer sep=0,baseline=(img.south)]
        \setlength{\picwidth}{.47\linewidth}
        \node[anchor=south west,filledletter= {\large b}{yellow}{black}{0.5}](img){\includegraphics[keepaspectratio,width=\picwidth]{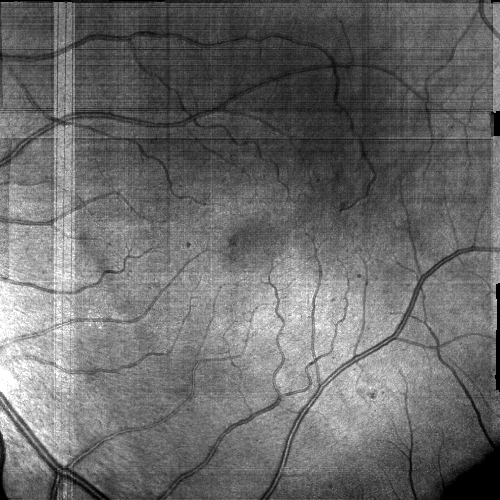}};
        \draw[green!70!black,ultra thick,loosely dashed,rotate=50](.9\picwidth,0\picwidth) ellipse (1.2cm and 1.4cm);
        \node[above left=.25cm and .25cm of img.south east, anchor=south east, draw=softblue,ultra thick](zoom){\includegraphics[keepaspectratio,width=.6\picwidth]{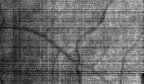}};
        \draw[softblue,ultra thick] ($(img.south west)+(.69\picwidth,.75\picwidth)$) rectangle ($(img.south west)+(.975\picwidth,.92\picwidth)$);
        \draw[softblue, ultra thick, dashed] (zoom.north west) -- (.69\picwidth,.75\picwidth);
        \draw[softblue, ultra thick, dashed] (zoom.north east) -- (.975\picwidth,.75\picwidth);
        \draw[<-,>=Latex,ultra thick,softred] (0.0\picwidth,.40\picwidth) -- ++(0,-.2cm);
        \draw[<-,>=Latex,ultra thick,softred] (0.0\picwidth,.75\picwidth) -- ++(0,.2cm);
        \draw[<-,>=Latex,ultra thick,softred] (.13\picwidth,.05\picwidth) -- ++(0,-.2cm);
        \draw[<-,>=Latex,ultra thick,softred] (.13\picwidth,.95\picwidth) -- ++(0,.2cm);
        \draw[<-,>=Latex,ultra thick,softred] (1.0\picwidth,.05\picwidth) -- ++(0,-.2cm);
        \draw[<-,>=Latex,ultra thick,softred] (1.0\picwidth,.95\picwidth) -- ++(0,.2cm);
    \end{tikzpicture}
    \caption{Global illumination variation (ellipse) and banding artifacts (arrows) in enface images (depth-averaged volumes). Left: Input with vertical B-scan orientation. Right: forward-warped \& merged.}
    \label{fig:scanning-artifact}
\end{figure}
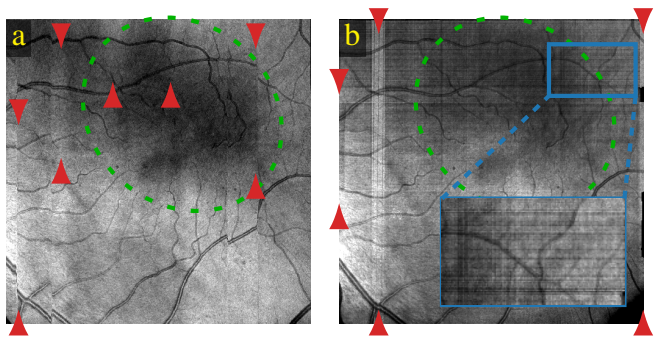

Retrospective motion correction generally reduces varying illumination via averaging, but can also introduce additional artifacts:
When image registration maps non-\-sub\-se\-quent\-ly acquired B-scans next to each other to correct distortions, only spatial continuity can be restored, while temporal continuity is not. Similar to scan discontinuities, this can appear as a banding-like artifact.
This especially applies to larger, saccadic motion that is responsible for quilting artifacts.
In these cases, forward warping motion correction alone, which only corrects distortion, is insufficient to remove the artifact.
Lastly, the coregistered volumes usually differ in illumination, and the merged data displays the average illumination. However, if one of the registered volumes has a gap, merged image intensity abruptly changes to the illumination level of the remaining volume(s), which can again result in a banding-like artifact.
The same effect occurs when eye-blinks are removed from a scan before merging.
Similarly, in forward warping, the merged volume is not computed by uniform averaging between the coregistered volumes. In contrast, the distance from the merged voxel to each individual A-scan can be taken into account by assigning closer A-scans a higher weight. While this can be advantageous for spatial accuracy, it makes the contribution from each volume change on a per-voxel basis, and with it, interpolates between the illumination levels of the merged volumes.

While banding artifacts are distracting for human readers (\figurename~\ref{fig:scanning-artifact}a), they can confound algorithmic quantitative disease markers: sharp changes can cause false edges in segmentation of vessels or pathologic features, and change in intensity can adversely affect thresholding as used in the segmentation of
retinal layers, or hyperreflective foci.
The illumination variation in forward-warped and merged images appears as intensity deviation that follows the B-scan acquisition directions, which can make them look similar to quilting artifacts (\figurename~\ref{fig:scanning-artifact}b). However, given the continuity of the underlying illumination, many artifact lines appear over larger areas, which can be prohibitive, especially in clinical use of forward warping motion correction, for both human and algorithmic image analysis.

There are only few studies on the correction of artifacts from inter-scan illumination differences in OCT, and general illumination normalization approaches can not be transferred because they do not model the anisotropy in OCT.
In the work of Wang et al., stitched OCTA volumes were linearly interpolated over a wide range to achieve smooth transitions between illumination levels \cite{Wang2019}.
However, wide range interpolation cannot compensate sharp illumination changes between scan lines or per voxel weight changes in forward warping.
%
Kraus et al.\ reported an unsharp mask-like illumination normalization method as preprocessing step for motion correction \cite{Kraus2014}.
However, by normalization, the approach alters morphology of image features like the darker appearing optic nerve head (ONH), or geographic atrophy (GA), a pathology appearing in late-stage age-related macular degeneration. Removal of such features would be detrimental to clinical image analysis.
Furthermore, the approach does not take the anisotropy of illumination into account, leaving residual artifacts at sharp changes.

In this paper, we propose an inverse model formulation of illumination bias which not only models the spatial, but, for the first time, also temporal (dis)continuity of illumination on a per-A-scan level.
By fitting the model via minimization of inter-scan illumination differences in registered OCT volumes, strong changes of illumination, as they can appear between B-scans of a single scan, and merging artifacts from illumination differences between scans, are resolved. At the same time, contrast of morphology like the foveal avascular zone or the ONH, and pathologies like GA, are preserved.
By using scans with orthogonal B-scan directions, sharp anisotropic changes in illumination are accurately estimated and corrected.


\section{Methods}
\label{sec:methods}

We will first define the illumination model and proceed with its optimization, which is performed by iterative image reconstruction.
For prior displacement field computation, and efficient warping of OCT images, we refer to our recent work on motion correction \cite{Ploner2022}.

\subsection{Illumination Model and Parameterization}
\label{ssec:model}

We model the measured signal $t$ as the product of illumination strength $a$ with sample backscattering $t^*$, and added noise $n$.
\begin{equation}
	t = a \cdot t^* + n \label{eq:illu}
\end{equation}
For strong backscattering, the term is dominated by the multiplicative part which can, given knowledge of $a$, be inverted to obtain an estimate $t^* \approx \frac 1a \cdot t$. For weak backscattering (in the background), the noise term dominates and $t^*$ cannot be estimted.
In these cases, a correction with $\frac 1a$ should not be applied to avoid noise amplification. More importantly, these signals must be excluded from the illumination estimation process to avoid bias. To maintain the appearance of OCT data and because the goal of this paper is limited to illumination artifact correction, we simply copy $t$, i.e., the noise-measurement, which is independent from $a$, and reconstruct the image
\begin{equation}
    \tilde t = \begin{cases}
        \frac 1a \cdot t & t^* \text{ in foreground} \\
        t & t^* \text{ in background}
    \end{cases} = \left(\frac 1a\right)^m \cdot t\label{eq:inv}
\end{equation}
To decide between the cases, we perform a first estimate of $t^*$ by median-filtering the input B-scans, to reduce noise, and apply a small threshold $t_\text{min}$. The result is a binary foreground mask-value $m$.
Since deviations in illumination are ignored, this estimation is inaccurate for large $t^*$, but the threshold value is chosen at the transition to the background noise level, where the multiplicative part has small influence (because $t^*$ is small and $a$ is generally close to $1$).

To simplify computations and increase numeric stability, we evaluate Eq.~\eqref{eq:inv} in logarithmic scale and define $s = \log t$.
Because A-scans are acquired in a single acquisition, there is no illumination variation within them. Since each B-scan is acquired spatially and temporally continuous, we assume continuous illumination within them. Due to temporal discontinuity and potential motion, we refrain from inter-B-scan assumptions. Therefore, we model the correction coefficient $\log \frac1a$ by a Hermite spline $c_{i}^V(j)$ for each B-scan $i$ and volume $V$ and index with the A-scan index $j$ over the spline parameters $\vec c$.
We use a parameter density of 1 per millimeter.
%
Given a voxel index $k$, the corrected log-signal $\tilde s$ can then be computed via
\begin{equation}
    \tilde s_{i,j,k}^V(\vec c) = s_{i,j,k}^V + m_{i,j,k}^V \cdot c_i^V(j).
\end{equation}

\subsection{Model Optimization}
\label{ssec:optimization}
\figurename~\ref{fig:method} sketches the optimization goal, which is a key to understanding the model optimization.
We minimize the objective 
\begin{equation}
	\mathcal J(\vec c) = \sum\nolimits_{M \in \mathcal V} \mathcal D^M(\vec c) + \lambda \cdot \mathcal R(\vec c) \quad \text{s.t.} \quad \mathcal C(\vec c) = 0,
\end{equation}
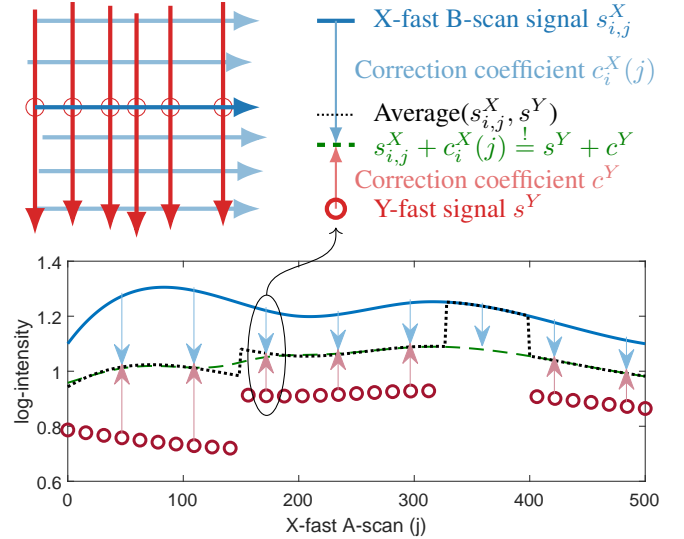
\begin{figure}[ht!]
    \centering
    \begin{tikzpicture}[outer sep=0,inner sep=0,baseline=1cm]
        \coordinate(grid) at (-2.5cm,0cm);
        \draw[->,>=Latex,ultra thick,softblue!50] ($(grid)+(-1.5cm+.00cm, 1.00cm-.00cm)$) -- ++(3cm, 0);
        \draw[->,>=Latex,ultra thick,softblue!50] ($(grid)+(-1.5cm-.10cm, 0.50cm-.05cm)$) -- ++(3cm, 0);
        \draw[->,>=Latex,ultra thick,softblue!50] ($(grid)+(-1.5cm+.10cm,-0.50cm-.05cm)$) -- ++(3cm, 0);
        \draw[->,>=Latex,ultra thick,softblue!50] ($(grid)+(-1.5cm+.05cm,-1.00cm+.00cm)$) -- ++(3cm, 0);
        \draw[->,>=Latex,ultra thick,softblue!50] ($(grid)+(-1.5cm+.00cm,-1.50cm-.00cm)$) -- ++(3cm, 0);

        \draw[->,>=Latex,ultra thick,softred] ($(grid)+(-1.5cm+.00cm,1.25cm-.10cm)$) -- ++(0, -3cm);
        \draw[->,>=Latex,ultra thick,softred] ($(grid)+(-1.0cm+.00cm,1.25cm+.00cm)$) -- ++(0, -3cm);
        \draw[->,>=Latex,ultra thick,softred] ($(grid)+(-0.5cm+.00cm,1.25cm-.05cm)$) -- ++(0, -3cm);
        \draw[->,>=Latex,ultra thick,softred] ($(grid)+( 0.0cm-.15cm,1.25cm-.15cm)$) -- ++(0, -3cm);
        \draw[->,>=Latex,ultra thick,softred] ($(grid)+( 0.5cm-.20cm,1.25cm-.05cm)$) -- ++(0, -3cm);
        \draw[->,>=Latex,ultra thick,softred] ($(grid)+( 1.0cm+.00cm,1.25cm+.00cm)$) -- ++(0, -3cm);

        \draw[softred] ($(grid)+(-1.5cm+.00cm,-.15cm)$) circle (.12cm);
        \draw[softred] ($(grid)+(-1.0cm+.00cm,-.15cm)$) circle (.12cm);
        \draw[softred] ($(grid)+(-0.5cm+.00cm,-.15cm)$) circle (.12cm);
        \draw[softred] ($(grid)+( 0.0cm-.15cm,-.15cm)$) circle (.12cm);
        \draw[softred] ($(grid)+( 0.5cm-.20cm,-.15cm)$) circle (.12cm);
        \draw[softred] ($(grid)+( 1.0cm+.00cm,-.15cm)$) circle (.12cm);

        \draw[->,>=Latex,ultra thick,softblue] ($(grid)+(-1.5cm+.00cm, 0.00cm-.15cm)$) -- ++(3cm, 0);
        
        \coordinate(ctr) at (0cm,-.65cm);
        \coordinate(avg) at (.25cm,-.25cm);
        
        \node[softblue,anchor=west] at (.5cm,1cm) {X-fast B-scan signal $s_{i,j}^X$};
        \draw[-,ultra thick,softblue] (-.25cm, 1cm) -- ++(.5cm, 0);
        
        \draw[->,>=Latex,thick,softblue!65] (0cm, 1cm) -- (ctr) node[pos=.4,anchor=west,outer xsep=.25cm]{Correction coefficient $c_i^X(j)$};
        
        \draw[-,thick,densely dotted] ($(avg)+(-.5cm,0)$) -- ++(.5cm, 0);
        \node[anchor=west,outer xsep=.25cm] at (avg){Average($s_{i,j}^X, s^Y$)};
        
        \draw[->,>=Latex,thick,softred!65] (0cm, -1.5cm) -- (ctr) node[midway,anchor=west,outer xsep=.25cm]{Correction coefficient $c^Y$};
        
        \draw[-,ultra thick,dashed,green!50!black] ($(ctr)+(-.25cm, 0cm)$) -- ++(.5cm, 0);
        \node[anchor=west,outer xsep=.5cm] at (ctr) {\color{green!50!black} $s_{i,j}^X + c_i^X(j) \stackrel{!}{=} s^Y + c^Y$};
        
        \draw[softred,ultra thick] (0cm,-1.50cm) circle (.12cm);
        \node[softred,anchor=west] at (.5cm,-1.50cm) {Y-fast signal $s^Y$};
        
        \node at (0,-4cm){\includegraphics[keepaspectratio,width=\linewidth]{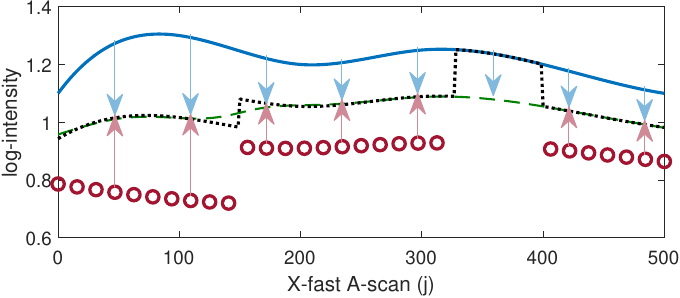}};
        \draw (-.92cm,-3.45cm) ellipse[x radius=.25cm, y radius=.8cm];
        \draw[->] (-.92cm,-2.65cm) to[out=90,in=-90] (0, -1.85cm) -- ++(0,.1cm);
    \end{tikzpicture}
    \caption{Compensation of illumination artifacts.
    Top left: Schematic of X-fast B-scans (blue) and Y-fast B-scans (red) after motion correction.
    Bottom: Visualization of the optimization goal in logarithmic scale, with legend / zoom in the top right.
    Blue curve: X-fast B-scan signal corresponding to the dark blue line in the top left.
    For clarity, we assume a sample with uniform (foreground) backscattering such that changes in measured signal only originate from illumination variance. Therefore, given the continuous illumination assumption in B-scans, the blue curve is continuous.
    The red circles represent the intensities of the orthogonal scan, which are interpolated and therefore written without indices. Because they originate from different B-scans, their illumination is not necessarily continuous.
    The dotted black curve, corresponding to the average of the uncorrected intensities, is discontinuous where $s^Y$ is discontinuous or has gaps. In forward warping, the averaging weights of the red intensities vary between points, which would distribute the black dots discontinuously somewhere between the blue and red inputs.
    The blue arrows represent the illumination correction of the X-fast intensities, which are continuous due to their spline parametrization.
    Added to the blue curve, they form the dashed green curve, the corrected X-fast intensities.
    Correction coefficients of the orthogonal scan are optimized such that the corrected intensities match the green line up to noise (in a different addend of the outer sum in Eq.~5).}
    \label{fig:method}
\end{figure}

where the constraint $\mathcal C(\vec c) = \sum c_i$ enforces global intensity to match the inputs. L$_2$-norm penalization $\mathcal R(\vec c) = ||\vec c||_2^2$ of illumination change preserves the original contrast, because it centers the illumination corrected result (green line) between the inputs. The dataterm 
\begin{equation}
	\mathcal D^M(\vec c) = \sum_{\substack{T \in \mathcal V \\ \dir(T) \\ \neq \dir(M)}} \sum_{i,j,k} \left( \tilde s_{i,j,k}^M(\vec c) - \mathcal W\big(\tilde S^T, \vec{\tilde x}_{i,j,k}^M \big) \right)^2
\end{equation}
penalizes the squared difference between each voxel (inner sum) of the {\it current} volume $M$ and each orthogonally scanned volume $T$ (outer sum). For this purpose, illumination corrected target volumes $\tilde S^T$ are Hermite-interpolated ($\mathcal W$) to the {\it current} voxel location $\tilde x_{i,j,k}^M$. If there is a gap in the target volume, the difference is excluded.\footnote{In contrast to \figurename~\ref{fig:method}, backscattering of non-constant samples varies arbitrarily depending on location. However, given correct co-registration, this location-dependent effect cancels out in the difference computation, making it irrelevant for optimization. Therefore, \figurename~\ref{fig:method} is a valid visualization of the optimization; and the continuity of correction coefficients along B-scans generalizes beyond the constant value assumption.}
We start the optimization by assuming uniform illumination ($\vec c := \vec 0$) and minimize via iterative momentum gradient descent.

The crux of the approach is that the inter-B-scan illumination continuity is enforced by the orthogonal target B-scans (whose spline-corrected illumination is guaranteed to remain continuous, \figurename~\ref{fig:method}). 
This also utilizes the registration to enforce illumination continuity in motion corrected space, as opposed to scanner space as in \cite{Kraus2014}, achieving continuous illumination in the merged image. Furthermore, gaps and varying merging weights are resolved: Because the corrected local intensities differ only by noise, weight changes / exclusion of a scan no longer impact the merged average.

\section{Evaluation \& Results}
\label{sec:evaluation}

\subsection{Data}

This research study was conducted retrospectively using human subject data. OCT imaging was performed at New England Eye Center (NEEC) in line with the principles of the Declaration of Helsinki.
Approval was granted by Committee on the Use of Humans as Experimental Subjects at MIT (\#1901669061) and Tufts Health Sciences Institutional Review Board at NEEC (\#MODCR-03-13240).
The dataset contains 18 subjects (70 $\pm$ 11 years old) with ocular pathologies, described in detail in \cite{Ploner2022}.
Four $6 \times 6$ mm fovea-centered scans were acquired with alternating B-scan orientation.
Illumination correction was performed on all 4 combinations of orthogonal scan pairs, totalling to 72 merged volumes. One subject was retrospectively excluded because illumination in the same peripheral region was so low in all 4 scans, that all retinal layers except the RPE vanished in background noise. A correction at this location is therefore impossible. Yet, other image areas in these scans were well corrected.
The prototype spectral domain OCT instrument images at 840 nm wavelength with 128 kHz A-scan rate, and acquired $500 \times 500$ A-scans with 3 \textmu m resolution (FWHM) \cite{ByungKun2020}. A-scans were reconstructed with 775 pixels and 1.78 \textmu m spacing.


\subsection{Qualitative Evaluation}

\begin{figure*}[tb]
    \centering
    \setlength{\picwidth}{.195\linewidth}
    \begin{tikzpicture}[inner sep=0, outer sep=0]
        \node(img)[filledletter={\large 1a}{yellow}{black}{0.5},anchor=south west] at (0, 0){\includegraphics[keepaspectratio,width=\picwidth]{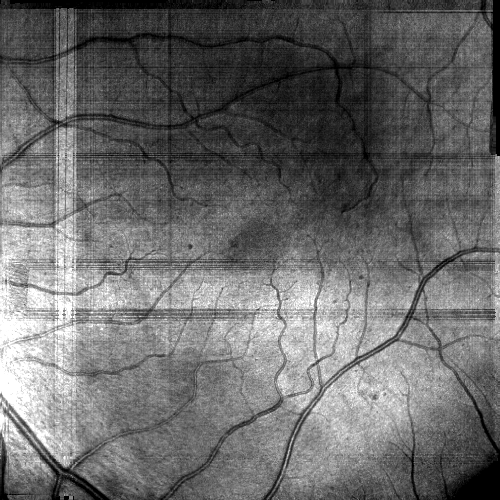}};
        \node[above left=.25cm and .25cm of img.south east, anchor=south east, draw=softblue,ultra thick](zoom){\includegraphics[keepaspectratio,width=.6\picwidth]{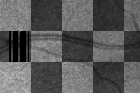}};
        \draw[softblue,ultra thick] ($(img.south west)+(.08\picwidth,.65\picwidth)$) rectangle ($(img.south west)+(.38\picwidth,.85\picwidth)$);
        \draw[softblue, ultra thick, dashed] (zoom.north west) -- (.08\picwidth,.65\picwidth);
        \draw[softblue, ultra thick, dashed] (zoom.north east) -- (.38\picwidth,.65\picwidth);
    \end{tikzpicture}
    \hfill
    \begin{tikzpicture}
        \node[inner sep=0, filledletter={\large 1b}{yellow}{black}{0.5},anchor=south west] at (0, 0){\includegraphics[keepaspectratio,width=\picwidth]{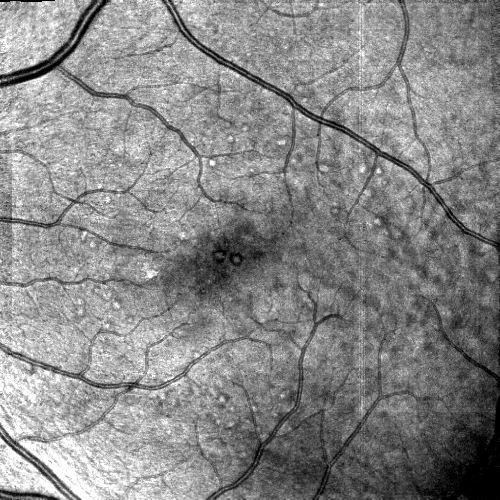}};
        \draw[<-,>=Latex,thick,softred] (.03\picwidth,.75\picwidth) -- ++(0,.2cm);
        \draw[<-,>=Latex,thick,softred] (.03\picwidth,.3\picwidth) -- ++(0,-.2cm);
        \draw[<-,>=Latex,thick,softred] (.725\picwidth,.935\picwidth) -- ++(0,.2cm);
        \draw[<-,>=Latex,thick,softred] (.725\picwidth,.15\picwidth) -- ++(0,-.2cm);
    \end{tikzpicture}
    \hfill
    \begin{tikzpicture}
        \node[inner sep=0, filledletter={\large 1c}{yellow}{black}{0.5},anchor=south west] at (0, 0){\includegraphics[keepaspectratio,width=\picwidth]{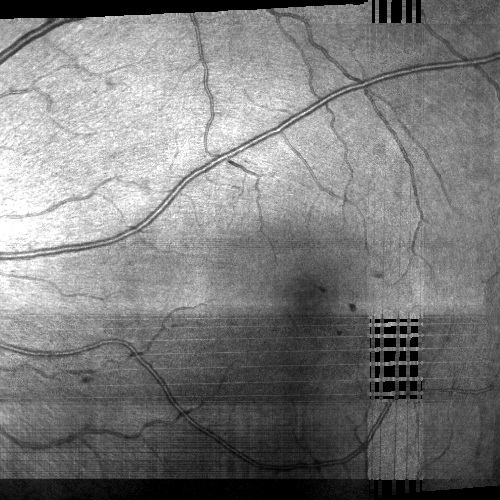}};
        \draw[<-,>=Latex,thick,softred] (.74\picwidth,.6\picwidth) -- ++(0,.2cm);
        \draw[<-,>=Latex,thick,softred] (.74\picwidth,.075\picwidth) -- ++(0,-.2cm);
        \draw[<-,>=Latex,thick,softred] (.845\picwidth,.6\picwidth) -- ++(0,.2cm);
        \draw[<-,>=Latex,thick,softred] (.845\picwidth,.075\picwidth) -- ++(0,-.2cm);
        \draw[->,>=Latex,thick,softred] (.015\picwidth,.2\picwidth) -- ++(.2cm,0);
        \draw[->,>=Latex,thick,softred] (.99\picwidth,.2\picwidth) -- ++(-.2cm,0);
        \draw[->,>=Latex,thick,softred] (.015\picwidth,.37\picwidth) -- ++(.2cm,0);
        \draw[->,>=Latex,thick,softred] (.99\picwidth,.37\picwidth) -- ++(-.2cm,0);
    \end{tikzpicture}
    \hfill
    \begin{tikzpicture}
        \node[inner sep=0, filledletter={\large 1d}{yellow}{black}{0.5},anchor=south west] at (0, 0){\includegraphics[keepaspectratio,width=\picwidth]{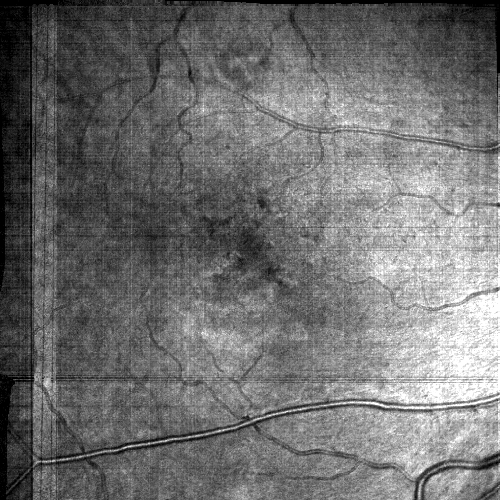}};
        \draw[<-,>=Latex,thick,softred] (.095\picwidth,.075\picwidth) -- ++(0,-.2cm);
    \end{tikzpicture}
    \hfill
    \tikz{\node[inner sep=0, filledletter={\large 1e}{yellow}{black}{0.5}]{\includegraphics[keepaspectratio,width=\picwidth]{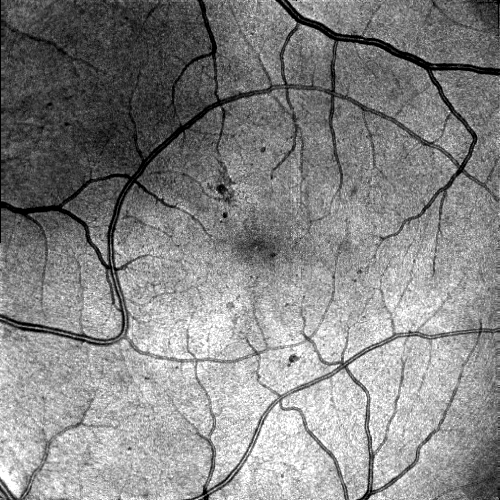}};}
    \\
    \begin{tikzpicture}[inner sep=0, outer sep=0]
        \node(img)[filledletter={\large 2a}{yellow}{black}{0.5},anchor=south west] at (0, 0){\includegraphics[keepaspectratio,width=\picwidth]{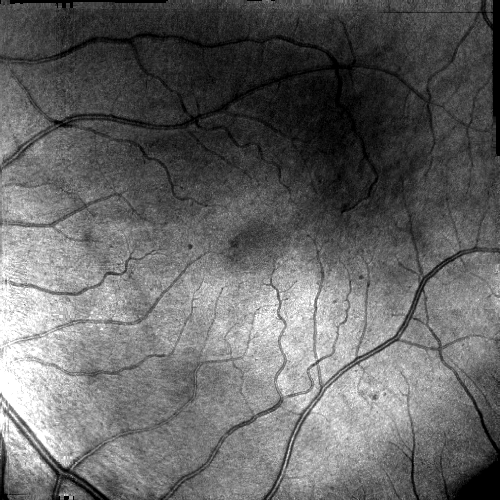}};
        \node[above left=.25cm and .25cm of img.south east, anchor=south east, draw=softblue,ultra thick](zoom){\includegraphics[keepaspectratio,width=.6\picwidth]{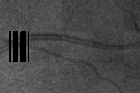}};
        \draw[softblue,ultra thick] ($(img.south west)+(.08\picwidth,.65\picwidth)$) rectangle ($(img.south west)+(.38\picwidth,.85\picwidth)$);
        \draw[softblue, ultra thick, dashed] (zoom.north west) -- (.08\picwidth,.65\picwidth);
        \draw[softblue, ultra thick, dashed] (zoom.north east) -- (.38\picwidth,.65\picwidth);
    \end{tikzpicture}
    \hfill
    \tikz{\node[inner sep=0, filledletter={\large 2b}{yellow}{black}{0.5}]{\includegraphics[keepaspectratio,width=\picwidth]{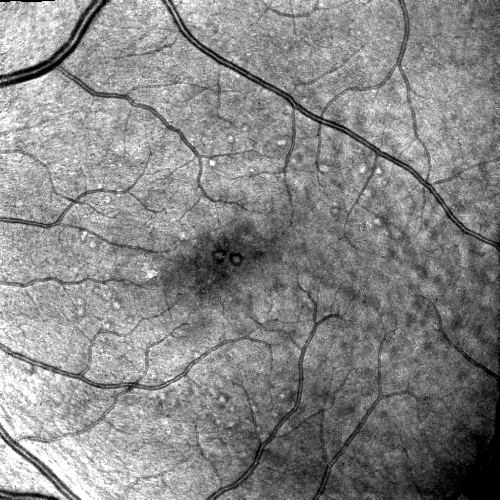}};}
    \hfill
    \begin{tikzpicture}
        \node[inner sep=0, filledletter={\large 2c}{yellow}{black}{0.5},anchor=south west] at (0, 0){\includegraphics[keepaspectratio,width=\picwidth]{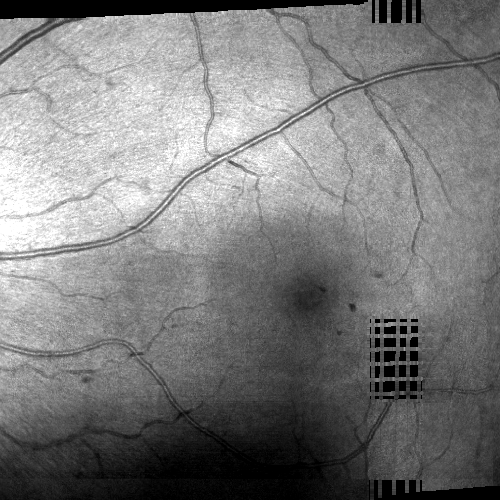}};
        \draw[<-,>=Latex,thick,softred] (.74\picwidth,.075\picwidth) -- ++(0,-.2cm);
        \draw[<-,>=Latex,thick,softred] (.845\picwidth,.075\picwidth) -- ++(0,-.2cm);
    \end{tikzpicture}
    \hfill
    \begin{tikzpicture}
        \node[inner sep=0, filledletter={\large 2d}{yellow}{black}{0.5},anchor=south west] at (0, 0){\includegraphics[keepaspectratio,width=\picwidth]{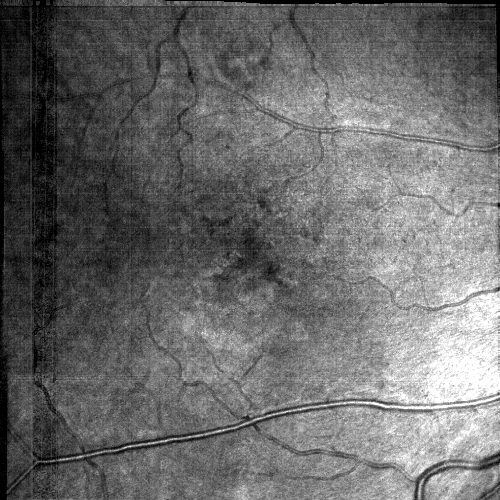}};
        \draw[<-,>=Latex,thick,softred] (.095\picwidth,.075\picwidth) -- ++(0,-.2cm);
    \end{tikzpicture}
    \hfill
    \begin{tikzpicture}
        \node[inner sep=0, filledletter={\large 2e}{yellow}{black}{0.5},anchor=south west] at (0, 0){\includegraphics[keepaspectratio,width=\picwidth]{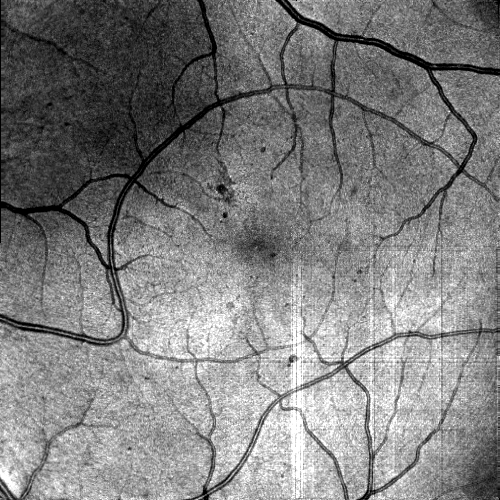}};
        \draw[<-,>=Latex,thick,softred] (.6\picwidth,.075\picwidth) -- ++(0,-.2cm);
        \draw[<-,>=Latex,thick,softred] (.753\picwidth,.075\picwidth) -- ++(0,-.2cm);
        \draw[<-,>=Latex,thick,softred] (.89\picwidth,.075\picwidth) -- ++(0,-.2cm);
    \end{tikzpicture}
    \caption{Representative enface images (depth-averaged volumes) from volumes merged without (row 1) and with illumination correction (row 2).
    (a) Good results after large and (b) small improvement. The zooms in (a) show a checkerboard overlay of the registered volumes before merging. Illumination difference makes the individual tiles in (1a) clearly distinguishable, while they are almost uniform after correction in (2a).
    (c) Minor residual banding artifacts (the black squares are gaps which originate from large saccadic motion).
    (d) Worst result (besides the residual banding artifact, slight lines with residual illumination deviation are visible throughout on a monitor).
    (e) The volume with most worsening (the other two worsened much less). In B-scans (which are not depth-averaged), the same change of illumination would be less visible due to higher noise levels.
    Contrast is adjusted jointly per column, corruptions in dark areas arise from the publication process.}
    \label{fig:results}
\end{figure*}
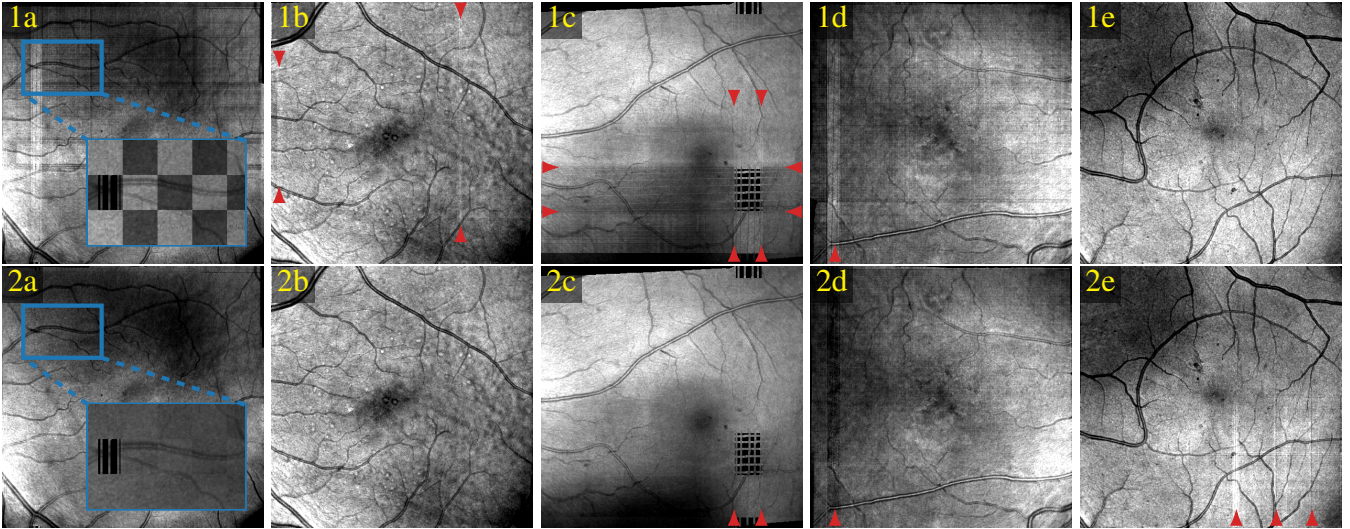

We categorized the result images by (a) severity of residual illumination artifacts and (b) change of illumination artifacts, and present the results in \tablename~\ref{tab:qualitative-results}. Minor artifacts are roughly evenly split among typically small-extent border artifacts, and more globally appearing but weak artifacts. The latter ones are barely visible in the enface images, and vanish in the higher noise present in non-averaged volumetric data. Representative examples are presented in \figurename~\ref{fig:results}.
\begin{table}[htb]
    \centering
    \begin{tabular}{c|c c c|c c}
        $\downarrow$Result \textbackslash{}\ change$\rightarrow$  &  + & o & - &    &      \\\hline
        no artifacts       & 45 & 2 & 0 & 47 & 69.1\% \\
        minor artifacts    & 14 & 2 & 1 & 17 & 25.0\% \\
        moderate artifacts &  1 & 1 & 2 &  4 &  5.9\% \\\hline
                           & 60 & 5 & 3 & 68 &      \\
                           & 88.2\% & 7.4\% & 4.4\% & \\
    \end{tabular}
    \caption{Contingency table of categories no residual artifacts, minor residual artifacts and moderate residual artifacts; and improvement (+), inconclusive change (o), and worsening (-).}
    \label{tab:qualitative-results}
\end{table}

\subsection{Quantitative Evaluation}

While the strong banding artifacts often result from gaps in one of the volumes, the line artifacts that show up in forward-warped data (\figurename~\ref{fig:scanning-artifact}b) originate from illumination difference between the registered volumes. To confirm our previous observations, we evaluated, if the illumination correction reduces the signal difference between the registered (but not merged) volumes. To reduce the impact of noise, we averaged the volumes along depth, where we only included registered voxels that were present in both volumes to avoid bias. Complying with the predominantly additive noise assumption, resampling and averaging was performed in linear scale. (Images in \figurename~\ref{fig:results} were computed in the same way). When computing the differences, we found a reduction in mean absolute difference of 22.5\%. Note that additional effects, like change in focus, misregistration, noise, and others, also increase the difference, so illumination correction alone can only reduce the difference, but not reach zero. 66 differences decreased, while 2 increased (which coincided with 2 of the 3 cases observed to have worsened in \tablename~\ref{tab:qualitative-results}, while the third one had the second-smallest decrease and was close to zero).

\section{Discussion}
\label{sec:discussion}

We performed an in-depth analysis on the formation of image artifacts in OCT and designed a physically-motivated model that can compensate for illumination variance. We parameterized the model using only the relevant degrees of freedom to achieve most-accurate compensation of temporal changes via overdetermined estimation, while the spline-based formulation ensures
that clinically relevant features are maintained.
By enforcing the corrected intensities to fit the average intensities, spatial illumination effects are averaged and therefore converge to zero with increasing number of scans.
Optimization was performed using the first 3D inverse model in OCT.

In our challenging clinical dataset, our approach reduces ill\-um\-in\-a\-tion-\-re\-la\-ted imaging artifacts in 88\% of cases, while only 4.4\% worsened.
After correction, only 6\% of cases had moderate residual illumination artifacts, which were easily identifiable in B-scan slices. We reconfirmed these numbers with a quantitative evaluation of residual difference between registered data after illumination correction.
Weak artifacts only become visible after axial averaging, but vanish in noise when viewing B-scans, which is the typical clinical use-case.
Small residual effects can sometimes be seen at the image boundary. While gaps inside volumes are not an issue for curve estimation, extrapolation at the image borders is more error-prone. More sophisticated extrapolation could potentially further reduce artifacts in these areas.
On a second level, this work enables use of forward warping-registration for accurate motion correction and merging of clinical data. Therefore, largely improved accuracy, resolution of stretch artifacts, and more reliable correction of doubling and quilting artifacts are enabled \cite{Ploner2022}.

Despite the promising results, our method has limitations:
Orthogonal scanning and accurate registration is a prerequisite.
Furthermore, image continuity and smallest correctable scale of illumination is a trade-off, therefore shadows from small-scale floaters are not corrected. However, since floaters are moving, they would be reduced by volume averaging.

By largely resolving the multiplicative illumination effect, this work enables high-quality 3D denoising via iterative reconstruction methods. Furthermore, super-resolution reconstruction of volumes and enface images is enabled including the transverse directions. Here, due to the denser sampling in the merged volume, individual voxels are only merged from a subset of the input scans, while neighboring pixels are filled from other scans that differ in illumination. Scan merging weights and, consequently, averaged brightness would vary from pixel to pixel.
As illumination inconsistency increases with scan time, its correction will remain relevant in widefield imaging and in OCT angiography scan protocols \cite{Spaide2018}, especially if multiple repeats are required for flow speed analysis \cite{Ploner2016}.

\section{Author Contributions} {SP is the main author. Algorithm design, evaluation, implementations, initial writing: SP. Data: JW, JG, KL, NW. Funding: SP, AM, JF. All authors
discussed the results and their presentation, and contributed to the final manuscript.}

\section{Acknowledgment}
\label{sec:acknowledgment}

We acknowledge funding by the Deutsche Forschungsgemeinschaft (DFG, German Research Foundation), project 508075009, and National Institutes of Health, project {5-R01-EY011289-36}.

\bibliographystyle{IEEEbib}
\bibliography{refs}

\begin{thebibliography}{10}

\bibitem{Huang1991}
David Huang, Eric~A. Swanson, and Charles~P. Lin,
\newblock ``Optical coherence tomography,''
\newblock {\em Science}, vol. 254, no. 5035, pp. 1178--1181, 11 1991.

\bibitem{Fujimoto2016}
James Fujimoto and Eric Swanson,
\newblock ``The development, commercialization, and impact of optical coherence tomography,''
\newblock {\em {Invest. Ophthalmol. Vis. Sci.}}, vol. 57, no. 9, pp. OCT1, 2016.

\bibitem{Spaide2018}
RF~Spaide, JG~Fujimoto, NK~Waheed, SR~Sadda, and G~Staurenghi,
\newblock ``Optical coherence tomography angiography,''
\newblock {\em Progress in Retinal and Eye Research}, vol. 64, pp. 1--55, 2018.

\bibitem{Holmen2020}
Ian~C. Holmen, Sri~Meghana Konda, and Jeong~W. Pak,
\newblock ``{Prevalence and Severity of Artifacts in Optical Coherence Tomographic Angiograms},''
\newblock {\em JAMA Ophthalmology}, vol. 138, no. 2, pp. 119--126, 02 2020.

\bibitem{Anvari2021}
Pasha Anvari, Maryam Ashrafkhorasani, and Abbas Habibi,
\newblock ``Artifacts in optical coherence tomography angiography,''
\newblock {\em J. Ophthal. {\&} Vis. Research}, vol. 16, no. 2, pp. 271--286, 2021.

\bibitem{Brea2019}
Luisa Sánchez~Brea, Danilo Andrade De~Jesus, and Muhammad~Faizan Shirazi,
\newblock ``Review on retrospective procedures to correct retinal motion artefacts in {OCT} imaging,''
\newblock {\em Applied Sciences}, vol. 9, no. 13, 2019.

\bibitem{Ploner2021}
Stefan~B. Ploner, Martin~F. Kraus, and Eric~M. Moult,
\newblock ``Efficient and high accuracy 3-d oct angiography motion correction in pathology,''
\newblock {\em Biomed. Opt. Express}, vol. 12, no. 1, pp. 125--146, Jan 2021.

\bibitem{Camino2016}
Acner Camino, Miao Zhang, and Simon~S. Gao,
\newblock ``Evaluation of artifact reduction in oct angiography with real-time tracking and motion correction technology,''
\newblock {\em Biomed. Opt. Express}, vol. 7, no. 10, pp. 3905--3915, Oct 2016.

\bibitem{Ploner2022}
Stefan Ploner, Siyu Chen, and Jungeun Won,
\newblock ``A spatiotemporal model for precise and efficient fully-automatic 3d motion correction in oct,''
\newblock in {\em Medical Image Computing and Computer Assisted Intervention -- MICCAI 2022}, 2022, pp. 517--527.

\bibitem{Wang2019}
Jie Wang, Acner Camino, Xiaohui Hua, Liang Liu, David Huang, Thomas~S. Hwang, and Yali Jia,
\newblock ``Invariant features-based automated registration and montage for wide-field oct angiography,''
\newblock {\em Biomed. Opt. Express}, vol. 10, no. 1, pp. 120--136, Jan 2019.

\bibitem{Kraus2014}
Martin Kraus, Jonathan Liu, and Julia Schottenhamml,
\newblock ``Quantitative {3D-OCT} motion correction with tilt and illumination correction, robust similarity measure and regularization,''
\newblock {\em {Biomed. Opt. Express}}, vol. 5, no. 8, pp. 2591--2613, 7 2014.

\bibitem{ByungKun2020}
ByungKun Lee, Siyu Chen, and Eric~M. Moult,
\newblock ``{High-Speed, Ultrahigh-Resolution Spectral-Domain OCT with Extended Imaging Range Using Reference Arm Length Matching},''
\newblock {\em Trans. Vis. Sci. \& Tech.}, vol. 9, no. 7, pp. 12--12, 06 2020.

\bibitem{Ploner2016}
Stefan~B. Ploner, Eric~M. Moult, and WooJhon Choi,
\newblock ``Toward quantitative optical coherence tomography angiography,''
\newblock {\em Retina}, vol. 36, pp. S118--S126, 2016.

\end{thebibliography}

\end{document}